\documentclass{article}
\usepackage{ijcai26}

\usepackage{times}
\usepackage{soul}
\usepackage{url}
\usepackage[hidelinks]{hyperref}
\usepackage[utf8]{inputenc}
\usepackage[small]{caption}
\usepackage{graphicx}
\usepackage{amsmath}
\usepackage{amsthm}
\usepackage{amssymb}
\usepackage{booktabs}
\usepackage{algorithm}
\usepackage[switch]{lineno}
\usepackage{subcaption}
\usepackage{algpseudocode} 
\usepackage{diagbox}
\usepackage{multirow}
\usepackage{xcolor}

\title{Rethinking Convolutional Networks for Attribute-Aware \\ Sequential Recommendation}

\author{
Shereen Elsayed$^{*,1,2}$,
Ngoc Son Le$^{*,1,2}$,
Ahmed Rashed$^{3}$
{\normalfont and}
Lars Schmidt-Thieme$^{1,2}$\\
\affiliations
$^1$Information Systems and Machine Learning Lab (ISMLL), University of Hildesheim, Germany\\
$^2$VWFS Data Analytics Research Center (VWFS DARC), University of Hildesheim, Germany\\
$^3$Volkswagen Financial Services AG, Germany\\
\emails{
\{elsayed,sle,schmidt-thieme\}@ismll.de,
ahmed.galal.ahmed.rashed@vwfs.com}
}

\begin{document}
\maketitle
\begin{abstract}
Attribute-aware sequential recommendation entails predicting the next item a user will interact with based on a chronologically ordered history of past interactions, enriched with item attributes. Existing methods typically leverage self-attention mechanisms to aggregate the entire sequence into a unified representation used for next-item prediction. While effective, these models often suffer from high computational complexity and memory consumption, limiting their ability to process long user histories. This constraint restricts the model's capacity to fully capture long-term user preferences. In some scenarios, modeling item interactions purely through attention may also not be the most effective approach to extract sequential patterns.
In this work, we propose \textbf{ConvRec}, an alternative method with linear computational and memory complexity that employs convolutional layers in a hierarchical, down-scaled fashion to generate compact, yet expressive sequence representations. To further enhance the model's ability to capture diverse sequential patterns, each layer aggregates the neighboring items gradually to reach a comprehensive sequence representation. 
Extensive experiments on four real-world datasets demonstrate that our approach outperforms state-of-the-art sequential recommendation models, highlighting the potential of convolution-based architectures for efficient and effective sequence modeling in recommendation systems. Our implementation code and datasets are available at \url{https://github.com/ismll-research/ConvRec}.

\end{abstract}

\begingroup
\renewcommand{\thefootnote}{\fnsymbol{footnote}}
\footnotetext[1]{Both authors contributed equally to this work.}
\footnotetext{Accepted at IJCAI-ECAI 2026.}
\endgroup





\section{Introduction}

\noindent
In the era of explosive growth of social media networks, video streaming, and e-commerce platforms, recommendation systems have become an indispensable tool to help users navigate the vast digital landscape by providing personalized suggestions that align with their unique preferences and past behaviors. More recently, sequential recommendation has emerged as a dominant recommendation task. Unlike traditional collaborative filtering (CF) methods, such as Matrix Factorization (MF) \cite{koren2009matrix,rendle2012bpr} that solely focus on static user-item interactions, sequential recommendation models are designed to capture the temporal dynamics of user behaviors. 
\begin{figure}[t]
\centering
\includegraphics[width=\columnwidth]{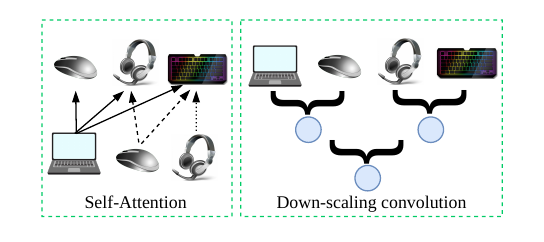} 
\captionof{figure}{Illustration of the self-attention versus the down-scaling convolution mechanism.}
\label{fig:conv-mechanism}
\end{figure}

 Earlier approaches in sequential recommendation domain were based on Markov Chains \cite{shani2005mdp}, \cite{rendle2010factorizing}, which could capture short-term dependencies but struggled with longer-range patterns. With the arrival of deep learning, a new wave of model architectures that are capable of handling high-order, nonlinear, and long-range data sequences was introduced. Models using Recurrent Neural Networks (RNNs) \cite{cho2014learning}, \cite{hidasi2018recurrent} could effectively capture longer evolving sequential dependencies, but suffer from training and inference inefficiency due to the lack of parallelization. On the other hand, Transformer-based models managed to address these limitations by leveraging self-attention to model dependencies across all items in a sequence at the same time, which eliminates the need for sequential processing and enables full parallelization during training. Models, such as SASRec \cite{kang2018self}, BERT4Rec \cite{sun2019bert4rec}, helped establish self-attention as a dominant modeling paradigm in the field of sequential recommendation. Despite parallelization capabilities, the computational and memory complexity of self-attention scales quadratically with the sequence length, which offsets much of the training efficiency it provides.

 In this paper, we revisit Convolutional Neural Networks (CNNs) as a competitive framework for temporal and attribute-aware sequential recommendation. Although early CNN-based models like Caser \cite{tang2018personalized} demonstrated the ability to capture local transition patterns, its receptive field is strictly bounded within the fixed history window of the last few items. Thus, it is unable to capture the global view of the long-range sequential dependencies in the way Transformers can. To overcome this limitation, we propose a hierarchical convolutional encoder that progressively downsamples the entire interaction sequence (via striding/pooling) to a single node, thereby distilling the temporal features into a compact "intent" representation (Figure \ref{fig:conv-mechanism}) with a global view of the sequence. This design takes inspiration from the pyramidal and patch-based techniques that were successfully used in time-series and vision models \cite{wang2024timemixer++}, \cite{lin2017feature}. 

 We introduce \textbf{ConvRec}, a hierarchical down-scaling convolutional model for attribute-aware sequential recommendation. First, it fuses each item embedding in the user's interaction history with its corresponding attribute embeddings and contextual embeddings into a joint representation. These embeddings form a chronological interaction sequence for each user, which is fed through a series of convolutional blocks. In each block, a strided convolution operation aggregates local neighboring interactions and downsamples the sequence. By repeating this down-scaling procedure in a pyramidal fashion, ConvRec progressively distills the entire user interaction history into a single vector with rich representation of the entire sequence, which is then used to score and rank next-item candidates via a dot-product operation.

Our primary contributions are as follows:
\begin{itemize}
    \item We propose a novel convolutional encoder for sequential recommendation, which leverages item attributes and context, and applying a hierarchical down-scaling convolutions to capture both local and long-range user interaction patterns. 
    \item Experiments across widely used sequential recommendation datasets, demonstrating superior performance of ConvRec in comparison to state-of-the-art baselines.
    \item We show that ConvRec scales linearly in memory and computation time with sequence length, and offers substantial efficiency improvements over existing best performing Transformer-based models on long user sequences.
\end{itemize}

\section{Related Work}
\textbf{Sequential Recommendation.} Sequential recommendation aims to predict the next item a user will interact with and is central to modern recommender systems. Earlier models like Caser \cite{tang2018personalized}, CosRec \cite{yan2019cosrec}, and RCNN \cite{xu2019recurrent} used convolutions and recurrent layers to capture sequential patterns. GRU4Rec \cite{jannach2017recurrent} and BERT4Rec \cite{sun2019bert4rec} introduced recurrent and transformer architectures, while SASRec \cite{kang2018self} employed multi-head self-attention. Extensions such as TiSASRec \cite{li2020time}, CoSeRec \cite{liu2021contrastive}, and TiCoSeRec \cite{dang2023ticoserec} incorporate temporal and contrastive learning signals to further improve performance.
\textbf{Attribute-aware Sequential Recommendation.} Incorporating item attributes is essential for capturing user preferences more accurately. $S^3$Rec \cite{zhou2020s3} leverages pre-training to enhance data representations. Other models, such as NOVA \cite{liu2021noninvasive}, DIFSR \cite{xie2022decoupled}, and MSSR \cite{lin2024multi}, propose various strategies for feature fusion. CARCA \cite{rashed2022context} introduces cross-attention mechanisms to compute item relevance scores more effectively. ProxyRCA \cite{seol2024proxy} builds upon CARCA by incorporating proxy-based item embeddings, significantly improving representation quality. Additionally, HGTL \cite{xu2025heterogeneous} applies cross-domain transfer learning to leverage user preferences from multiple domains.

Despite these advancements, convolutional architectures remain relatively underexplored in the context of sequential recommendation, even though they have shown strong potential for capturing local patterns in sequences \cite{liu2022scinet,wang2024timemixer++}. In this work, we propose a lightweight, hierarchical convolutional approach for down-scaling and encoding user sequences, offering an efficient and effective alternative for sequence representation.


\section{Problem Setting}
In the attribute-aware sequential recommendation setting, we consider a set of users $\mathcal{U}$ interacting with a set of items $\mathcal{I}$. Each user $u \in \mathcal{U}$  has a chronologically ordered interaction sequence $S^u = [i_1^u, \ldots, i_{|S^u|}^u]$, where each item is associated with a set of descriptive features/attributes $A$ (e.g., title, brand, etc.), the items features sequence for user $u$ is denoted as $[a_1^u, \ldots, a_{|S^u|}^u]$. Additionally, each interaction is accompanied by contextual information $C$, such as the timestamp of the interaction, which can be decomposed into temporal components (e.g., year, month, day). We denote the contextual sequence for user $u$ as $[c_1^u, \ldots, c_{|S^u|}^u]$. The primary objective of the next-item recommendation task is to predict the most likely item a user will interact with at the next time step.

\section{ConvRec}
Considering the similarity between time-series forecasting and next-item recommendation tasks, sequential recommendation can similarly benefit from the hierarchical scaling properties used in time series. Specifically, a user’s input sequence can be aggregated in a hierarchically down-scaled manner to derive a meaningful sequence representation. Given that neighboring items in a sequence are often more closely related than distant ones \cite{wang2020next,kang2018self}, unlike the self-attention mechanism, we propose a representation method where items are gradually merged with their neighbors to form a unified representation of the entire sequence.


\begin{figure}[t]
\centering
\includegraphics[width=\columnwidth]{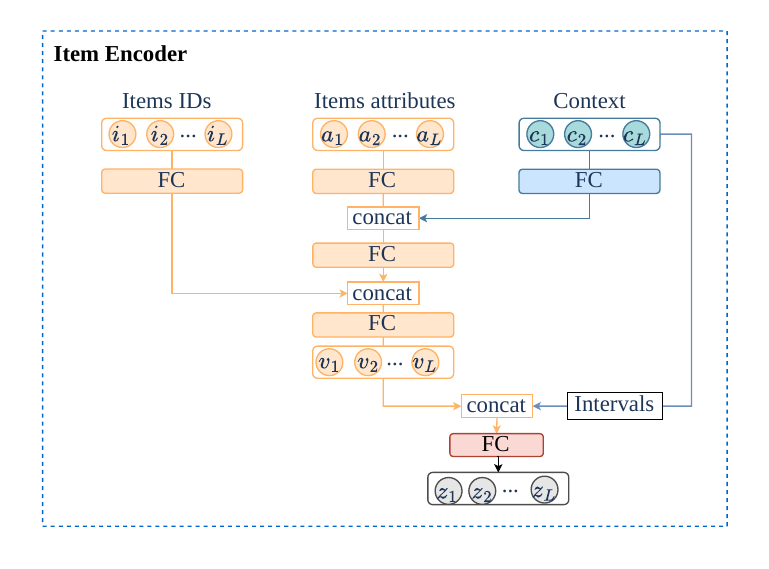}
\caption{Item encoding module.}
\label{fig:item-encoder}
\end{figure}

\subsection{Item Encoding}

For item encoding, we rely not only on item IDs, but also emphasize the importance of incorporating item attributes/features. These attributes provide insight into user preferences and enhance the model's ability to accurately predict the next item. For a given input sequence of fixed length $L$ (obtained through either padding or truncation), each item's features are passed through a fully connected layer:

\begin{equation}
    {a'^u_i} = a^u_i  W_a + b_a
\end{equation}
$W_a \in \mathbb{R}^{|A|\times d_a}$ is a weight matrix, $|A|$ is the length of the item's features set, $d_a$ is the embedding dimension of the input features and $b_a$ is the bias term. In addition to item features, we incorporate interaction context, such as time and location of interactions. In this work, we focus on temporal context, which is also encoded via a fully connected layer:


\begin{equation}
    {c'^u_i} =  c^u_i W_c+ b_c
\end{equation}

$W_c \in \mathbb{R}^{|C| \times d_c}$ is the weight matrix for the context features, $|C|$ is the number of contextual elements, $d_c$ is the context embedding dimension, and $b_c$ is the bias term. The encoded item features and context vectors are then concatenated and passed through another fully connected layer to generate the interaction embedding:

\begin{equation}
    {{f}_i}^u =  \text{Concat}({a'^u_i},{c'^u_i}) W_{f}+ b_{f}
\end{equation}


Here, $W_f \in \mathbb{R}^{(d_a + d_c) \times d_f}$, $d_f$ is the output embedding dimension, and $b_f$ is the bias term.

While various strategies exist for incorporating item IDs in recommendation systems, we adopt the approach introduced in \cite{seol2024proxy}, which assigns unique embeddings only to the most frequent items. Less frequent items are assigned a shared generic ID. This approach shifts the model’s learning focus toward high-frequency items while relying on feature-based representations for infrequent ones. Accordingly, we embed item IDs using a lookup table: ${i'_i}^u = \text{LookupTable}({i_i}^u)$. This embedding is concatenated with the previously obtained feature-context embedding and passed through a linear layer:
\begin{equation}
    {{v}_i}^u =  \text{Concat}({i'_i}^u,{{f}_i}^u) W_{v}+ b_{v}
\end{equation}
$W_{v} \in \mathbb{R}^{{({d_i}+{d_{f}}})\times {d_{v}}}$,  $d_{v}$ is the embedding dimension of the combined item IDs and side information, and $b_{v}$ is the bias term.

In sequential recommendation, temporal dynamics play a critical role. The time gap between interactions can vary significantly, so time not only contributes to the representation of individual interactions but can also help assess their relative importance. To this end, we compute time intervals (year, month, day) between each two consecutive interactions, and concatenate the intervals to the input sequence, where for the first interaction time interval is set to zero:

\begin{align} 
   \Delta \mathbf{c}^u = \text{Concat}(\Delta \mathbf{c}^{0}, [ \mathbf{c}^{u}_{2:L} - \mathbf{c}^{u}_{1:L-1}]),\ \ \ \ \\ {Q}^u =  \text{LayerNorm}( \sigma(\text{Concat}({V}^u, {\Delta \mathbf{c}}^u)))
    \label{weight} 
\end{align}

Here, $c$ is the interaction context where we extract only the \textbf{(year, month and day)} for the interval calculation, ${V}^u = [v_1, v_2, \ldots, v_L]$ is the sequence of item representations, and $\Delta \mathbf{c^u} = [\Delta c^u_1, \Delta c^u_2, \ldots, \Delta c^u_L]$ are the corresponding intervals, and $\sigma$ denotes the GELU activation function \cite{hendrycks2016gaussian}. Then it is fed into a fully connected layer to obtain a combined representation.
\begin{equation}
    {Z}^u =  \sigma({Q}^u W_Q+ b_Q)
\label{enc}
\end{equation}
$W_Q \in \mathbb{R}^{{(d_{v}+3)} \times d_{v}}$ is a weight matrix, and $b_Q$ is the bias term, and $\sigma$ denotes the GELU activation function.

\begin{figure*}[!t]
    \centering
    \includegraphics[scale=0.78]{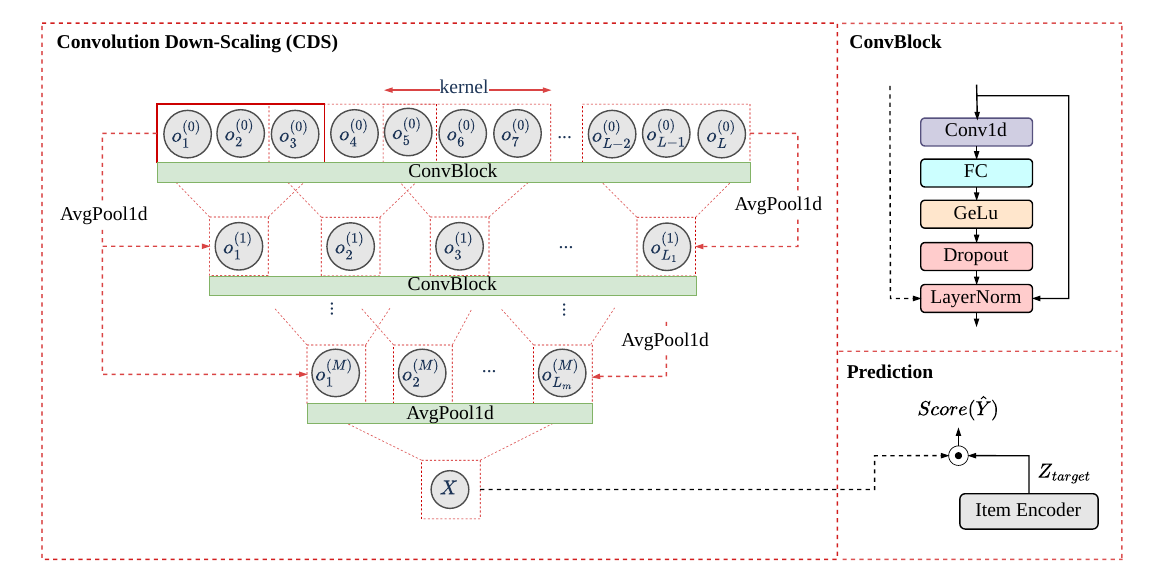}
    \caption{The ConvRec framework.}
    \label{fig:enter-label}
\end{figure*}

\subsection{Convolution Down-scaling Component (CDS)}

Building on the idea of gradual aggregation within a user's historical interaction sequence, we design a hierarchy of convolutional layers to construct the final sequence representation. Given an encoded sequence ${Z}^u \in \mathbb{R}^{L \times d_v}$, where $L$ denotes the maximum sequence length and $d_v$ is the dimensionality of the interaction embeddings, our objective is to reduce the sequence length $L$ to 1. This is achieved by progressively aggregating information through a series of hierarchical convolutional downscaling layers that downscale the sequence over multiple ConvBlocks: 
\begin{equation} 
\begin{aligned}
{X}^u &= \textbf{CDS}(Z^u) \\ 
&= \text{ConvBlock}_M \circ\dots\circ\text{ConvBlock}_1(Z^u)
\end{aligned}
\end{equation}
Here, $Z^u$ denotes the input embeddings, initially shaped as $L \times d_v$, where $L$ is the sequence length and $d_v$ is the embedding dimension, and $X^u$ denotes the output representation. To enable convolutional operations across neighboring items, the embeddings are reshaped such that the convolutional kernel can operate along the sequence dimension with $d_v$ input channels.\\

\textbf{Convolution Block (ConvBlock).} 
To perform the convolution operations effectively, the input sequence $Z^u$ is transposed to shape $d_v \times L$, treating $d_v$ as the number of input channels. This reshaping allows 1D convolution to be applied along the sequence dimension $L$:

\begin{equation} 
\begin{aligned}
&{K^{(j)}}^u = \text{1D-Conv}(\text{Padding}({{O^{(j-1)}}^u}^T),  \text{KernelSize}, \text{Stride} ),
 \\ 
&\textbf{where } \; {O^{(0)}}^u= Z^u
\end{aligned}
\label{eq:9}
\end{equation}
The 1D convolution is applied across the sequence dimension ($L$) using a specified kernel size and stride. In most cases, the stride is set equal to the kernel size to ensure non-overlapping windows,  which empirically yielded better performance. When the sequence length is not divisible by the kernel size, appropriate padding is applied to maintain dimensional consistency. After convolution, ${K^{(j)}}^u$ is transposed back to the shape $L^{(j)} \times d_v$, where $L^{(j)}$ is the sequence length after applying the convolution operation. A final fully connected layer is then applied:
\begin{equation} 
{G^{(j)}}^u = \sigma\left({{K^{(j)}}^u}^T W_G + b_G\right)
\label{eq:10}
\end{equation}
Here, $W_G \in \mathbb{R}^{{d_{K^{(j)}}} \times {d_{G^{(j)}}}}$ is the weight matrix of the output projection layer, $b_G$ is the bias term, and $d_G$ is the output embedding dimension, which is also typically set equal to $d_v$ for consistency and $\sigma$ is a GELU activation.

Subsequently, layer normalization is performed on the output. Additionally, we incorporate two types of weighted residual connections to enhance training stability:


\begin{equation}
{\small
\begin{aligned}
{O^{(j)}}^u = \text{LayerNorm}\bigg(
{G^{(j)}}^u &+ \alpha_1 \cdot \text{AvgPool}(Z^u) \\
&+ \alpha_2 \cdot \text{ProgRes}^{(j-1)}
\bigg)
\end{aligned}
}
\label{eq:11}
\end{equation}
The first residual connection is derived by applying average pooling to the original input. This helps preserve the original signal and mitigates the risk of vanishing gradients. The second, $\text{ProgRes}^{(j)}$, is a progressive residual connection from the previous convolutional layer's output. It is also downsampled using average pooling to match the output dimensions.

In the first convolutional layer, both residual connections originate from the same input, as no prior layer exists. However, in subsequent layers, $\text{ProgRes}$ continues to evolve, allowing for smoother gradient flow and improved model convergence. $\alpha_1$ and $\alpha_2$ are learnable parameters that are initialized to 0.5.



Consequently, the downscaling convolutional blocks can be defined as follows:

\begin{equation}
\begin{aligned}
&\textbf{for } j = 1,\dots,M \textbf{ do:} \\
&\quad {O^{(j)}}^{u}
= \text{ConvBlock}_{j}(
\text{Padding} ({O^{(j-1)}}^{u}),
\text{KernelSize},\text{Stride}), \\
&\quad \textbf{where } \; {O^{(0)}}^{u} = Z^{u}.
\end{aligned}
\label{eq:12}
\end{equation}

Here, $M$ denotes the total number of convolutional layers, resulting in the final output ${O^{(M)}}^u$ with an expected shape of $d_v \times 1$. However, in cases where the convolutional layers do not completely collapse (using average pooling), the sequence length is reduced to \textbf{1}. An additional average pooling layer is applied to ensure the final output has the desired dimension.

\begin{equation} 
X^u = \text{Collapse}({O^{(M)}}^u) 
\label{eq:13}
\end{equation}

In this step, ${O^{(M)}}^u$ is the output of shape $1 \times d_v$.

\begin{algorithm}
\caption{Forward pass of the ConvRec model}\label{alg:convrec}
\begin{algorithmic}[1]
\Require input encoded sequence ${Z}^u$, list of layer specs $M = [(kernel_1,stride_1),\dots,(kernel_m,stride_m)]$
\Require activation function $\sigma(\cdot)$, pooling function $\text{Pool}(\cdot)$
\Ensure logits $\hat{Y}$
\vspace{4pt}

\State ${ProgRes}^{(0)} \gets Z^u$ 
\State ${O^{(0)}}^u \gets Z^u$ 

\For{$j = 1 \to M$}
        \State ${{K^{(j)}}^u} \gets  \text{1D-Conv}(\text{Padding}({{O^{(j-1)}}^u}^T), kernel_j,$
        \Statex \hspace{1.5em} $\text{stride}_j )$ (Eq: \ref{eq:9})

        \State ${G^{(j)}}^u \gets \sigma({{K^{(j)}}^u}^T W_G + b_G)$ (Eq: \ref{eq:10})
        
        \State ${O^{(j)}}^u \gets \text{LayerNorm}({G^{(j)}}^u + \alpha_1 \cdot \text{AvgPool}(Z^u) + \alpha_2 \cdot \text{ProgRes}^{(j-1)}) $ (Eq: \ref{eq:11})
        
        \State ${ProgRes}^{(j)} \gets {O^{(j)}}^u$ 
\EndFor

\State $X^u \gets \text{Collapse}({O^{(M)}}^u)$ (Eq: \ref{eq:13})
\State $\hat{Y} \gets  Z^{u}_{target} \cdot (X^u)^T $ (Eq: \ref{eq:14})
\State \Return $\hat{Y}$
\end{algorithmic}
\end{algorithm}

\subsection{Model Training and Optimization}
For item ranking, we calculate the final score by taking the dot product of  the \textbf{target items' embedding} $Z^{u}_{target}$ obtained from the item encoder (Equation \ref{enc}) and the transpose of the sequence embedding $(X^u)^T$ as follows:
\begin{equation}
    \hat{Y} =  Z^{u}_{target} \cdot (X^u)^T 
    \label{eq:14}
\end{equation}

Given a positive item sampled from a user’s sequence and a set of negative items drawn from a candidate pool (excluding items in the user’s sequence), we train the model using the binary cross-entropy loss defined as follows:
\begin{equation}
      \mathcal{L}=  - \sum_{S \in \mathcal{S}}\left[  log{(\hat{Y}^{{(+)}})} +   \sum_{n=0} ^{N_{train}} (log{(1-\hat{Y}^{{(-)}}_{n})})\right]
       \label{eq:15`}
\end{equation}
\noindent where $\hat{Y}^{{(+)}}$ are the output scores for the positive sample and $\hat{Y}^{{(-)}}_{n}$ are the output scores for $N_{train}$ negative samples in our case we set $N_{train}$ to 100. $\mathcal{S}$ is the set of all sequences.

Many recommender models select the most recent part of the input sequence to represent the input and the shifted sequence to represent the positive items of the user \cite{kang2018self}, \cite{rashed2022context}. While effective, this approach narrowly focuses on recent interactions, potentially overlooking valuable information present in earlier parts of the user's history. In contrast, we adopt the training strategy proposed in \cite{seol2024proxy}, where instead of always selecting the latest subsequence, a random subsequence of length $L$ is sampled from the full user sequence $S^u$, and the final item in that sampled subsequence is used as the positive target. This allows the model to learn from various segments of the user’s behavior, not just the most recent interactions, and promotes better generalization.


\section{Experiments}

In this section we aim to answer the following research questions:
\begin{itemize}
    \item RQ1: What is the model performance versus state-of-the-art methods?
    \item RQ2: How effective is each model component?
    \item RQ3: What is the effect of varying model hyperparameters such as kernel sizes and strides? 

    \item RQ4: What are the computation and memory costs of the proposed model versus attention-based approaches? 
    
\end{itemize}

\textbf{Datasets.} To comprehensively evaluate our proposed model, we conduct experiments on four widely used Amazon datasets. The Fashion and Men datasets \cite{elsayed2022end}, \cite{rashed2022context}, \cite{seol2024proxy} are derived from different segments of the Amazon fashion domain. The Fashion dataset includes six categories: men's and women's tops, bottoms, and shoes. In contrast, the Amazon Men dataset comprises a broader range of subcategories, such as gloves, scarves, and sunglasses. For both datasets, item attributes are derived from visual features extracted using a pre-trained ResNet-50 model.
In addition, we include the Amazon Games dataset \cite{rashed2022context}, \cite{seol2024proxy}, which contains video game items along with attributes such as category and brand. Lastly, the Amazon Beauty dataset \cite{rashed2022context}, \cite{seol2024proxy} includes all items in the beauty category, where item attributes are also based on category and brand information. Table \ref{data} summarizes the statistics of the datasets.\\

\begin{table}[h]
\scalebox{0.7}{%
\centering
\begin{tabular}  {c cccc cc}
\hline
Dataset & Users & Items & Interactions& Avg. Seq. & Max. Seq. & Attributes\\ 
\hline
Beauty & 52,204 & 57,289 & 394,908 & 7 & 293 &  6507 \\
Games & 31,013 & 23,715 & 287,107 & 9 & 860 & 506 \\
Fashion & 45,184 & 166,270 & 358,003 & 8 &  192 &  2048 \\
Men & 34,244 & 110,636 & 254,870 & 7 & 308 & 2048 \\

\hline 
\end{tabular}
}
\caption{Datasets statistics}
\label{data}
\end{table}

\begin{table*}[h]
\centering
\setlength{\tabcolsep}{4pt}
\scalebox{0.75}{%
\begin{tabular}{c cc @{\hspace{15pt}} cc @{\hspace{15pt}} cc @{\hspace{15pt}} cc}
\hline
\multirow{2}{*}{Method}
& \multicolumn{2}{c}{Beauty}
& \multicolumn{2}{c}{Games}
& \multicolumn{2}{c}{Fashion}
& \multicolumn{2}{c}{Men} \\
\cmidrule(lr){2-3}
\cmidrule(lr){4-5}
\cmidrule(lr){6-7}
\cmidrule(lr){8-9}
& HR@10 & NDCG@10
& HR@10 & NDCG@10
& HR@10 & NDCG@10
& HR@10 & NDCG@10 \\
\hline

\textbf{BERT4Rec} & 0.478 & 0.318 & 0.705 & 0.509 & 0.328 & 0.209 & 0.315 & 0.193 \\
\textbf{SASRec} & 0.485 & 0.322 & 0.742 & 0.541 & 0.381 & 0.245 & 0.397 & 0.259 \\
\textbf{TiSASRec} & 0.492 & 0.333 & 0.748 & 0.533 & 0.384 & 0.234 & 0.333 & 0.194 \\
\textbf{CosRec} & 0.403 & 0.253 & 0.662 & 0.444 & 0.274 & 0.150 & 0.294 & 0.164 \\

\hline
\hline

\textbf{$S^3$-Rec} & 0.538 & 0.371 & 0.765 & 0.549 & 0.367 & 0.239 & 0.365 & 0.238 \\
\textbf{SASRec++} & 0.557 & 0.383 & 0.752 & 0.533 & 0.546 & 0.344 & 0.500 & 0.315 \\
\textbf{CARCA} & 0.579 & 0.396 & 0.782 & 0.573 & 0.591 & 0.381 & 0.550 & 0.349 \\
\textbf{CARCA-LIP} & 0.608 & 0.423 & 0.762 & 0.560 & 0.648 & 0.427 & \color{blue}{0.614} & 0.398 \\
\textbf{ProxyRCA} & \color{blue}{0.609} & \color{blue}{0.430}
& \color{blue}{0.805} & \color{blue}{0.601}
& \color{blue}{0.661} & \color{blue}{0.435}
& 0.613 & \color{blue}{0.399} \\

\hline

\textbf{ConvRec (ours)}
& \color{red}{0.629 $\pm$ {\tiny $1e{-}3$}}
& \color{red}{0.439 $\pm$ {\tiny $3.5e{-}3$}}
& \color{red}{0.819 $\pm$ {\tiny $1e{-}3$}}
& \color{red}{0.617 $\pm$ {\tiny $2.5e{-}3$}}
& \color{red}{0.688 $\pm$ {\tiny $2.6e{-}3$}}
& \color{red}{0.469 $\pm$ {\tiny $3.8e{-}3$}}
& \color{red}{0.618 $\pm$ {\tiny $1.2e{-}3$}}
& \color{red}{0.406 $\pm$ {\tiny $1.3e{-}3$}} \\

Improv.(\%)
& 3.28\% & 2.09\%
& 1.74\% & 2.66\%
& 4.08\% & 7.82\%
& 0.65\% & 2.00\% \\

\hline
\end{tabular}
}
\caption{Model performance and comparison against baselines on four benchmark datasets using \textbf{100} negative test samples. The best results are reported in \color{red}{red} \color{black}{and the second best in} \color{blue}{blue}.}
\label{results}
\end{table*}

\begin{table}[h]
\centering
\setlength{\tabcolsep}{4pt}
\scalebox{0.75}{%
\begin{tabular}{c cc @{\hspace{15pt}} cc}
\hline
\multirow{2}{*}{Method}
& \multicolumn{2}{c}{Beauty}
& \multicolumn{2}{c}{Games} \\
\cmidrule(lr){2-3}
\cmidrule(lr){4-5}
& HR@10 & NDCG@10
& HR@10 & NDCG@10 \\
\hline

\textbf{BERT4Rec} & 0.013 & 0.007 & 0.059 & 0.030 \\
\textbf{SASRec} & 0.008 & 0.004 & 0.026 & 0.014 \\
\textbf{TiSASRec} & 0.048 & \color{red}{0.027} & 0.095 & 0.047 \\
\textbf{CosRec} & 0.023 & 0.011 & 0.069 & 0.035 \\

\hline
\hline

\textbf{SASRec++} & 0.027 & 0.013 & 0.075 & 0.037 \\
\textbf{DIF-SR} & \color{blue}{0.051} & \color{blue}{0.025} & 0.109 & 0.048 \\
\textbf{CARCA-LIP} & 0.037 & 0.017 & 0.116 & 0.064 \\
\textbf{ProxyRCA} & 0.049 & \color{blue}{0.025} & \color{blue}{0.136} & \color{blue}{0.075} \\

\hline

\textbf{ConvRec (ours)}
& \color{red}{0.052}
& \color{red}{0.027}
& \color{red}{0.141}
& \color{red}{0.081} \\

Improv.(\%)
& 1.9\% & 0.0\%
& 3.6\% & 8.0\% \\

\hline
\end{tabular}
}
\caption{Model performance and comparison against baselines on the Beauty and Games datasets using \textbf{all items} as negative test samples. The best results are reported in \color{red}{red} \color{black}{and the second best in} \color{blue}{blue}.}
\label{results2}
\end{table}

\textbf{Evaluation Protocol.} We adopt a widely-used evaluation protocol based on the leave-one-out strategy \cite{kang2018self}, \cite{rashed2022context}, \cite{seol2024proxy}. Specifically, for each user, the last interacted item is held out for testing, the second-to-last item is used for validation, and the remaining sequence is used for training. For a fair comparison, we adopt the same negative sampling strategy used in prior studies \cite{rashed2022context}, \cite{seol2024proxy}, in which 100 negative items are sampled for evaluation. 

We report standard top-$K$ metrics, Hit Rate (HR), and Normalized Discounted Cumulative Gain (NDCG) to effectively capture the performance of our model.

\textbf{Baselines.} We selected several state-of-the-art sequential and attribute-aware sequential recommendation models to assess the performance of our proposed method.
\begin{itemize}
    
    
    \item BERT4Rec \cite{sun2019bert4rec}: A sequential recommendation framework based on a bidirectional Transformer, trained to predict randomly masked items within the input sequence.
    
    \item SASRec \cite{kang2018self}: A self-attention-based model that learns item dependencies in the sequence to generate next-item.
    \item CosRec \cite{yan2019cosrec}: A 2D convolutional model that encodes item sequences to capture complex pairwise and high-order item interactions.
    
    \item TiSASRec \cite{li2020time}: An extension of SASRec that integrates temporal information (e.g., time intervals between interactions) into the attention mechanism.
    
    \item $S^3$-Rec \cite{zhou2020s3}: An attribute-aware sequential recommendation model that employs self-supervised pre-training to enhance item and user representations.
    
    \item SASRec++ \cite{rashed2022context}: An enhanced version of SASRec that incorporates item attribute information to improve recommendation accuracy.
    
    \item DIF-SR \cite{xie2022decoupled}: A state-of-the-art approach that injects item features into the attention layers rather than restricting them to the input layer.
    
    \item CARCA \cite{rashed2022context}: A context-aware recommendation method that employs cross-attention to compute item scores. 
    
    \item CARCA-LIP \cite{seol2024proxy}: A CARCA model variant proposed in \cite{seol2024proxy} introduces an improved training approach and enhanced item embeddings, resulting in better performance.
    
    \item ProxyRCA \cite{seol2024proxy}: An advancement over CARCA that introduces proxy-based item embeddings, further improving item representation quality.
    
    
\end{itemize}

\subsection{Model Performance Against Baselines}

Table \ref{results} reports the HR@10 and NDCG@10 of all methods on the four benchmark datasets, evaluated with 100 negative samples per user. For each dataset, the mean and standard deviation are reported based on three independent runs. The baseline models were trained and tuned according to the guidelines and hyperparameter settings in their original papers. ConvRec achieves the best results in both HR@10 and NDCG@10 metrics across all datasets. ProxyRCA is the strongest baseline and ranks second in terms of overall performance in most of the cases. Relative to ProxyRCA, ConvRec yields improvements ranging from 0.65\% to 4.08\% in HR@10 and from 2\% to 7.82\% in NDCG@10. In overall, attribute-aware models outperform sequence-only baselines. Among them, all recent top performers are attention-based models, whereas, to the best of our knowledge, ConvRec is the only convolution-based attribute-aware model. \\
\indent Table \ref{results2} compares model performance on the Beauty and Games datasets using the all-item evaluation setting. While baseline sequential models (e.g., BERT4Rec, SASRec) perform modestly, stronger baselines such as TiSASRec and ProxyRCA achieve competitive results. Our proposed ConvRec surpasses all existing methods, delivering the best HR@10 and NDCG@10 on both datasets. ConvRec achieves improvements of up to 3.6\% in HR@10 and 8.0\% in NDCG@10 over the best baseline, indicating its enhanced ability to model sequential dependencies and user–item interactions.
\begin{figure*}[t]
    \centering
    \begin{subfigure}[b]{0.46\textwidth}
        \includegraphics[width=\linewidth]{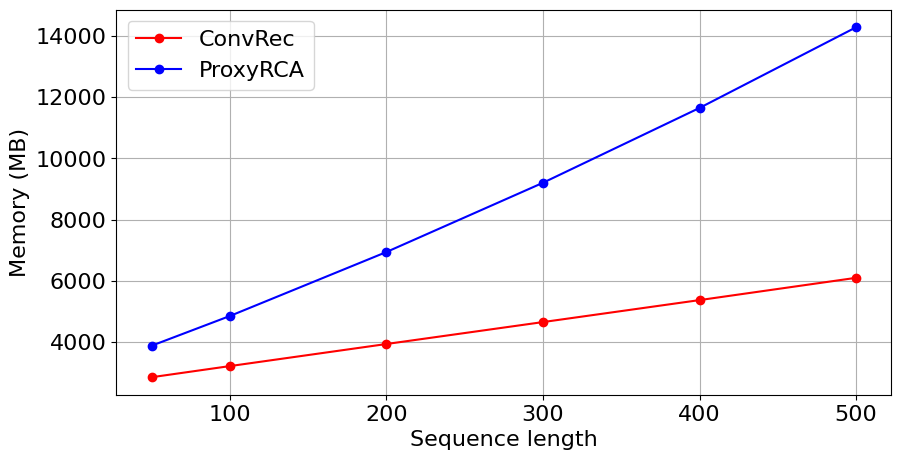}
        \label{fig:Mem}
    \end{subfigure} 
    \hfill 
    \begin{subfigure}[b]{0.46\textwidth}
        \includegraphics[width=\linewidth]{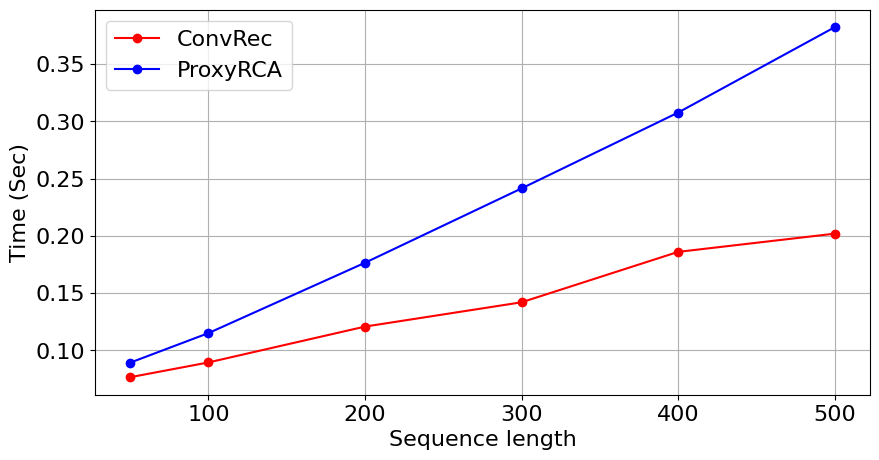}
        \label{fig:Time}
    \end{subfigure}
    \caption{Comparison of memory usage and execution time between ConvRec and ProxyRCA models at batch size 128.}
    \label{fig:mem}
\end{figure*}

\subsection{Ablation Studies}

The ablation studies are organized into four main parts:  
\begin{enumerate}
    \item \textbf{Analyzing the impact of individual model components} on overall performance.
    \item \textbf{Comparing memory and runtime efficiency} between the proposed convolution-based model and a multi-head attention-based model, such as ProxyRCA.
     \item \textbf{Exploring alternative convolutional architectures} by testing different kernel sizes and strides. Additionally, testing the hyperparameter sensitivity of the model. 
     
    \item \textbf{Evaluating the effect of varying sequence lengths} on model performance, using the \textit{Games} dataset, which has the longest average user sequence length.
   
\end{enumerate}

\begin{table}
\centering
\begin{tabular}  {c |cc cc}
\hline
   \diagbox{Study}{Dataset}  & Beauty & Games & Fashion & Men\\ 
\hline
w/o Intervals & 0.625 & 0.817 & 0.682 & 0.616  \\
w/o Residuals & 0.602 & 0.810 & 0.659 & 0.574  \\
w/ one Conv & 0.617 & 0.796 & 0.674 & 0.605  \\
w/ AvgPool & 0.602 & 0.804 & 0.669 &  0.611 \\
\hline
ConvRec & \textbf{0.629} & \textbf{0.819} & \textbf{0.688} &  \textbf{0.618} \\

\hline 
\end{tabular}
\caption{Ablation studies on different model components (HR@10)}
\label{ablation}
\end{table}

\subsubsection{Effect of Model Components}

Table~\ref{ablation} presents the impact of various model components on overall performance. One key factor examined is the concatenation of the time intervals between successive interactions. This component shows a particularly notable impact on the Beauty dataset, highlighting the importance of temporal information in modeling user behavior.

Another important component is the residual connection. Its effect is evident on all datasets for example in the Men dataset, removing the two residual connections leads to a significant performance drop from $0.618$ to $0.574$. This underscores the value of incorporating residual connections into the model architecture.

Furthermore, we evaluate a variant of the model that uses a single convolutional layer with both kernel size and stride equal to the sequence length. This configuration effectively acts as a fully connected layer over the entire input sequence. While its performance is slightly lower than the proposed model, the drop is not substantial.

Additionally, we consider a baseline that computes the average pooling of item embeddings without any additional layers. This approach generally results in a performance decline across most datasets. However, the drop is relatively minor in the Men dataset, where the performance decreases to $0.611$.

\begin{table}[h]
\centering
\begin{tabular}{c|cc}
\hline
(Kernel size, Stride) & HR@10 &    NDCG@10    \\
\hline
\textbf{\{ (2, 2), (5, 5), (7, 7) \}} & \textbf{0.630}
 & \textbf{0.438 }  \\
\{ (3, 3), (5, 5), (5, 5) \}  & 0.625  & 0.435   \\
\{ (5, 5), (2, 2), (7, 7) \}  & 0.624  &  0.435  \\
\{ (7, 7), (5, 5), (2, 2) \}  & 0.622  & 0.432   \\
\{ (10, 10), (7, 7) \} & 0.621  & 0.430   \\
\hline 
\{ (2, 1), (5, 5), (7, 7) \}  & 0.623  &  0.431 \\
\{ (2, 1), (5, 2), (7, 3) \}  & 0.615  & 0.421  \\
\{ (3, 1), (3, 1), (3, 1) \} & 0.593  & 0.399  \\
 \hline
\end{tabular}
\caption{Comparison between different kernel sizes, strides, and number of layers (Beauty dataset).}
\label{kernel}
\end{table}
\subsubsection{Memory and Runtime Efficiency}
We evaluated the training memory and runtime efficiency on Beauty dataset of our model and ProxyRCA using a batch size of 128 on NVIDIA A40 GPU. As shown in Figure \ref{fig:mem}, ProxyRCA exhibits a quadratic increase in memory consumption with respect to sequence length, rising from 3,878 MB for sequences of length 50 to 14,280 MB for sequences of length 500. In contrast, our model demonstrates linear memory growth, scaling from 2,849 MB to 6,096 MB under the same conditions.

Runtime performance follows a similar trend: ProxyRCA requires 0.0893 seconds per batch for sequences of length 50 and 0.382 seconds for length 500, while our model completes the same tasks in 0.076 and 0.202 seconds, respectively. These results highlight our model’s superior scalability in both memory efficiency and computational speed.

\subsubsection{Hyperparameter sensitivity } We explore alternative network architectures by experimenting with different kernel sizes. Furthermore, we analyze the impact of various hyperparameter settings, such as the embedding size and the dropout rate, to assess the model’s sensitivity to different configurations (Appendix A).  Lastly, we examined the performance across user groups with varying sequence lengths to evaluate the effectiveness of our convolution-based model compared to an attention-based model such as \textbf{ProxyRCA} (Appendix B).\\

\textit{Kernel size.} Since kernel size plays a crucial role in convolutional models, tuning it directly impacts the network architecture. As shown in Table~\ref{kernel}, we experimented with various kernel sizes across different numbers of layers. The choice of kernel size for the first layer proved particularly important for the final model performance. The best results were obtained with a first-layer kernel size of 2, where each pair of items is aggregated together, followed by subsequent layers that aggregate the next 5 groups, and so on. Additionally, we experimented with different strides, such as a stride of 1 to combine each item with its preceding and succeeding items for a richer representation. However, this did not improve performance, and using a stride equal to the kernel size yielded the best results.\\

\section{Conclusion}
This paper introduces ConvRec, a convolution-based model for attribute and context-aware sequential recommendation. By leveraging adaptive convolution layers, the model compresses user historical sequences into compact yet highly expressive representations. Experimental results demonstrate that ConvRec outperforms state-of-the-art methods in recommendation accuracy. Moreover, compared to traditional multi-head attention approaches, the proposed model achieves superior efficiency in both memory usage and runtime, making it scalable for real-world applications.

\newpage
\bibliographystyle{named}
\bibliography{ijcai26}

\clearpage
\newpage

\appendix

\section{Hyperparameter Sensitivity}
\label{Hyper}
\begin{figure}[h]
    \centering
    \begin{subfigure}[b]{0.49\textwidth}
        \includegraphics[width=\linewidth]{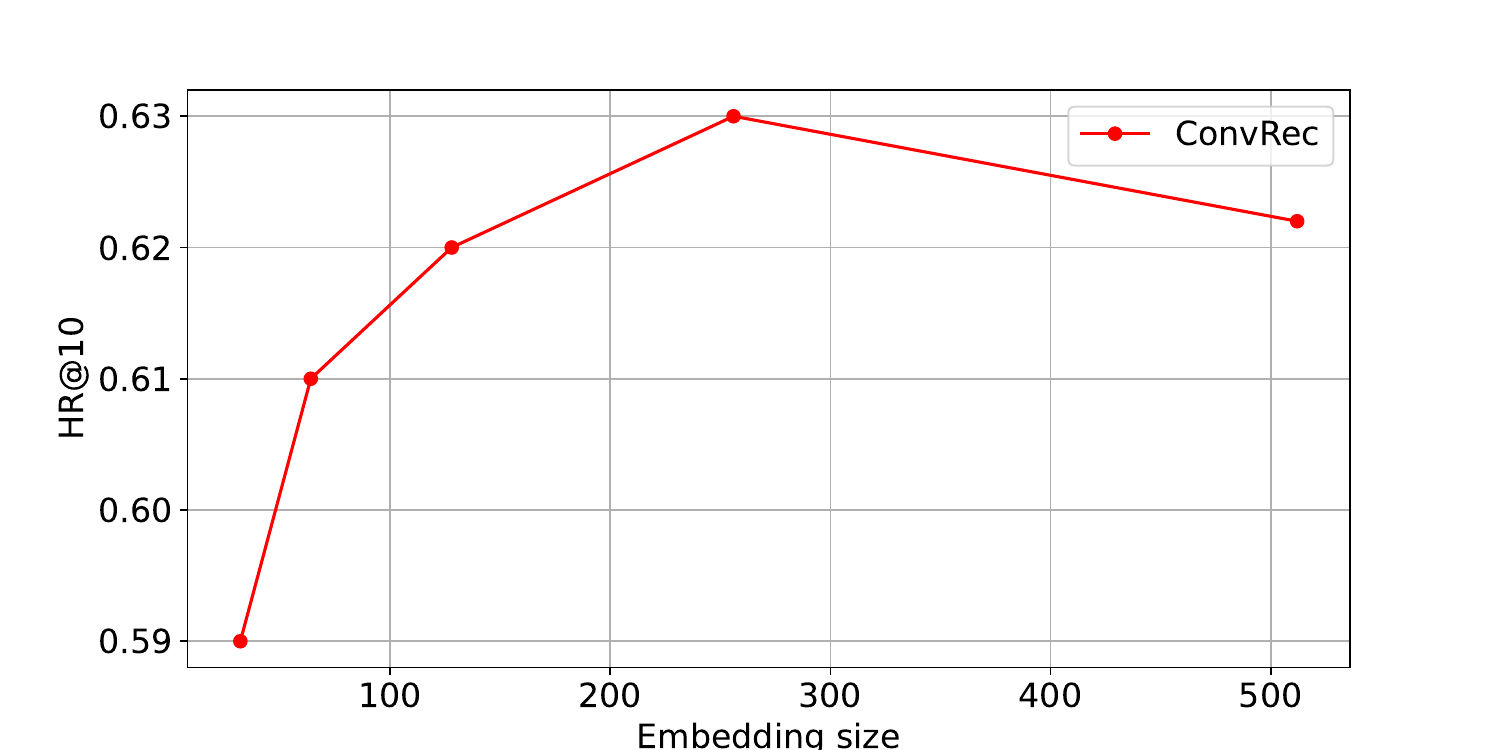}
        \label{fig:Emb}
    \end{subfigure} 
    \hfill 
    \begin{subfigure}[b]{0.49\textwidth}
        \includegraphics[width=\linewidth]{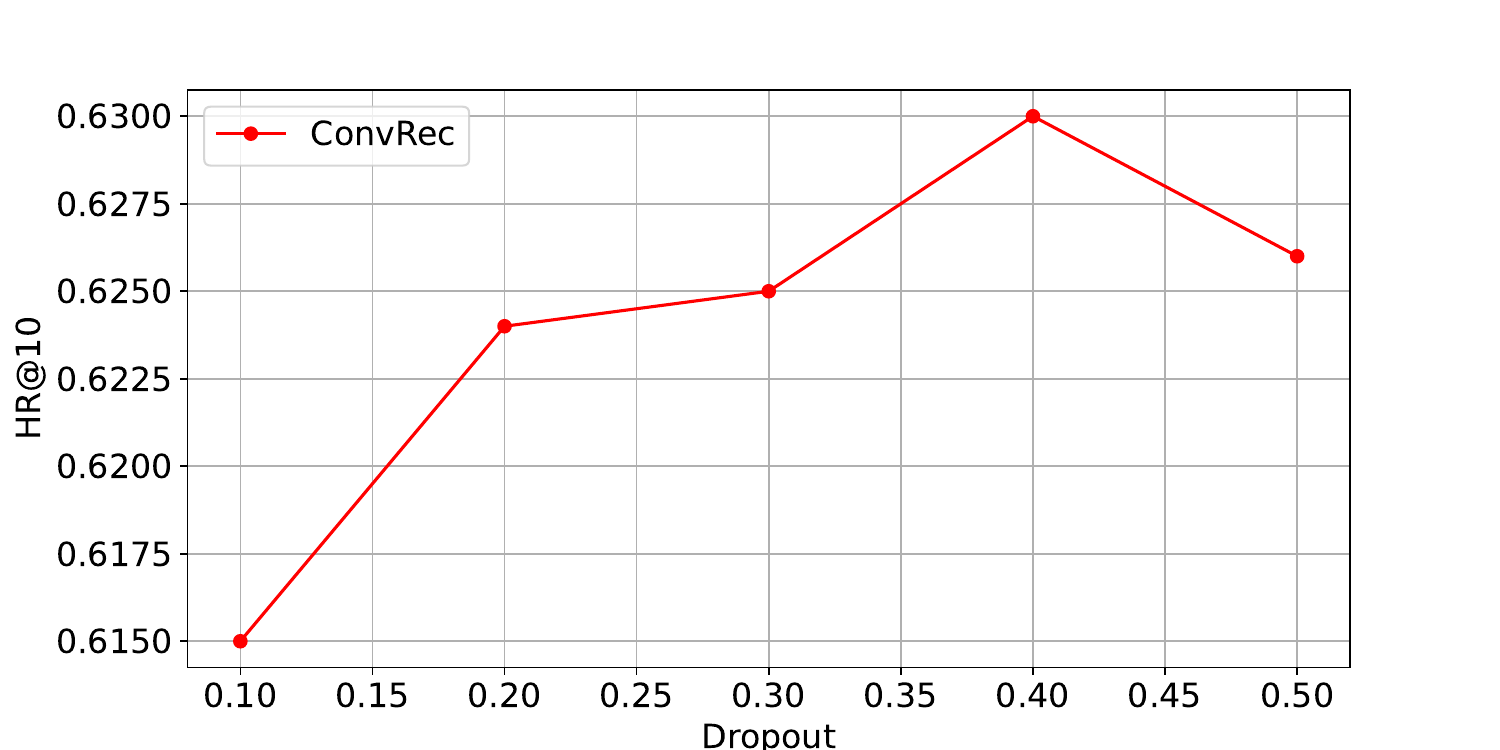}
        \label{fig:Drop}
    \end{subfigure}
    \caption{Effect of changing embedding size and dropout on model performance. }
    \label{fig:combined}
\end{figure}

\textit{Embedding Size and Dropout.} We evaluated different embedding sizes for item representation 32, 64, 128, 256, and 512 as shown in Figure \ref{fig:combined}. The smallest embedding size, \textbf{32}, yielded the weakest performance. Increasing the embedding size consistently improved results, reaching the best performance at an embedding size of 256. However, further increasing the embedding size beyond 256 led to a decline in performance. For dropout, we tested values of 0.1, 0.2, 0.3, 0.4, and 0.5. Increasing the dropout rate generally improved performance, with a value of 0.4 yielding the best results for the Beauty dataset, as shown in Figure \ref{fig:combined}. \\

\section{User-group Evaluation}
\label{Groups}
\noindent \textit{Varying sequence length.} Since the maximum sequence length significantly impacts the next-item prediction task, we selected the \textit{Games} dataset, characterized by the longest average user sequences. To evaluate the effect of varying input sequence lengths for both \textbf{ConvRec} and \textbf{ProxyRCA}. As illustrated in Table \ref{Top_Bottom}, \textbf{ConvRec} consistently outperforms \textbf{ProxyRCA} across all tested sequence lengths. Notably, a sequence length of 50 yields the best performance for both models.

Additionally, we divide users into two groups: the \textbf{Top} group, consisting of the top 20\% of users with the longest interaction sequences, and the \textbf{Bottom} group, comprising the remaining 80\% of users with shorter sequences. For both groups, we evaluate the performance of our proposed model, \textbf{ConvRec}, against the attention-based \textbf{ProxyRCA} model across varying sequence lengths. As shown in Table \ref{Top_Bottom}, the two models perform comparably within the \textbf{Top} group. However, in the \textbf{Bottom} group, \textbf{ConvRec} consistently outperforms \textbf{ProxyRCA}, demonstrating the effectiveness of the convolution-based approach for users with shorter sequences.
\begin{table}
\centering
\scalebox{0.7}{
\begin{tabular}{c|cc|cc|cc}
\hline
\multirow{2}{*}{Seq. length} & \multicolumn{2}{c|}{ Overall } & \multicolumn{2}{c|}{ Top 20\% } & \multicolumn{2}{c}{Bottom 80\% } \\
                               & ConvRec &  ProxyRCA  & ConvRec &  ProxyRCA  & ConvRec & ProxyRCA \\
\hline
10  & \textbf{0.811}  & 0.804  & \textbf{0.822}  &  0.816 & \textbf{0.807}  & 0.800  \\
20  & \textbf{0.812}  & 0.805  & 0.823  & \textbf{0.826} & \textbf{0.810}  &  0.799 \\
30  & \textbf{0.812}  & 0.805  & \textbf{0.826}  &  0.822 & \textbf{0.809}  &  0.800 \\
40  & \textbf{0.814}  & 0.804  & \textbf{0.827}  &  0.817 &  \textbf{0.811}  &   0.801 \\
50  & \textbf{0.815}  & 0.806  & 0.827  & \textbf{0.832}  &  \textbf{0.812} &  0.799 \\
100 & \textbf{0.812}  & 0.804  & \textbf{0.824}  & 0.814  & \textbf{0.808}  &  0.801 \\
200 & \textbf{0.813}  & 0.801  & \textbf{0.824}  & 0.805  & \textbf{0.810}  &  0.800 \\
300 & \textbf{0.814}  & 0.803  & \textbf{0.825}  & 0.820  & \textbf{0.810}  &  0.798 \\
\hline
\end{tabular}}
\caption{Performance comparison between ConvRec and ProxyRCA on different sequence lengths and on top and bottom users (HR@10/ Games dataset).}

\label{Top_Bottom}
\end{table}

\section{Hyperparameter Settings}



Our experiments were conducted on an NVIDIA A40 GPU using PyTorch \cite{paszke2019pytorch}. We performed grid-search tuning over embedding sizes (16–512), learning rates ($1\times10^{-5}$–$3\times10^{-4}$), sequence lengths (30–250), weight decay (0.05–0.3), and dropout (0.1–0.7). The best overall setup used a batch size of 128 and an embedding dimension of 256 across all datasets. Optimal learning rates were 0.00004 (Beauty), 0.0001 (Games), 0.00005 (Fashion), and 0.0001 (Men). Dropout rates were 0.45, 0.35, 0.35, and 0.3, with corresponding weight decays of 0.2, 0.1, 0.1, and 0.1. The best sequence lengths were 70 (Beauty), 50 (Games and Fashion), and 30 (Men). Models were trained for up to 1000 epochs with early stopping, using the Adam optimizer \cite{kingma2014adam}.

\end{document}